**Title:** Combating anti-statistical thinking using simulation-based methods throughout the undergraduate curriculum

**Authors:** Tintle N[1], Chance B[2], Cobb G[3], Roy S[4], Swanson T[5], VanderStoep J[6]

1. Associate Professor of Statistics, Department of Mathematics, statistics and Computer Science, Dordt College, Sioux Center, Iowa
2. Professor of Statistics, Department of Statistics, Cal Poly – San Luis Obispo, San Luis Obispo, CA
3. Robert L Rooke Professor, Department of Mathematics and Statistics, Mount Holyoke College, South Hadley, MA
4. Associate Professor of Statistics, Department of Statistics, Cal Poly – San Luis Obispo, San Luis Obispo, CA
5. Associate Professor of Mathematics, Department of Mathematics, Hope College, Holland, MI
6. Assistant Professor of Mathematics, Department of Mathematics, Hope College, Holland, MI

**Abstract:** The use of simulation-based methods for introducing inference is growing in popularity for the Stat 101 course, due in part to increasing evidence of the methods ability to improve students' statistical thinking. This impact comes from simulation-based methods (a) clearly presenting the overarching logic of inference, (b) strengthening ties between statistics and probability/mathematical concepts, (c) encouraging a focus on the entire research process, (d) facilitating student thinking about advanced statistical concepts, (e) allowing more time to explore, do, and talk about real research and messy data, and (f) acting as a firmer foundation on which to build statistical intuition. Thus, we argue that simulation-based inference should be an entry point to an undergraduate statistics program for all students, and that simulation-based inference should be used throughout all undergraduate statistics courses. In order to achieve this goal and fully recognize the benefits of simulation-based inference on the undergraduate statistics program we will need to break free of historical forces tying undergraduate statistics curricula to mathematics, consider radical and innovative new pedagogical approaches in our courses, fully implement assessment-driven content innovations, and embrace computation throughout the curriculum.

# 1. Introduction

For far too long much of our effort in statistics education has been limited by a narrow focus on the algebra-based introductory course ('Stat 101'). For the most part, that beginning course has been regarded as a terminal service course, and, unlike first courses in other STEM subjects, has offered neither an introduction to the undergraduate major in statistics, nor even a clear path to a second applied course in support of some other major (Johnstone, 2014). It has been very rare for students to progress from an applied introductory statistics course to a graduate program in statistics. Accordingly, in the context of the insular nature of Stat 101, this special issue of *TAS* is a welcome acknowledgement that it is time to turn our attention to courses in statistics beyond simply Stat 101, also considering goals and choices for a second or third applied course and identifying good models for innovative undergraduate courses in programs for minors and majors.

Despite the importance of courses beyond the first one, we also regard it as important not to sever our thinking about the introductory course for future majors from the rest of the statistics curriculum. Within the last decade, the algebra-based introductory course has been the focus of significant pedagogical and content reform efforts. In that spirit, this article describes the rationale behind the growing reform effort incorporating simulation and randomization-based methods for teaching inference in the introductory statistics course, with an eye towards the implications of this curricular reform throughout the undergraduate statistics curriculum.

In Section 2, we will lay out a framework for describing obstacles to developing statistical thinking. We will argue that there are two major, forces that hinder students from developing statistical thinking: (1) focusing courses and curriculum on mathematical (deductive) reasoning, and (2), failing to address the commonly-held, dismissive belief that statistics is overly pliable, and, thus, not reliable.

As we will describe in Section 3, the new reform movement of simulation-based introductory courses for non-majors (Stat 101) is one way to help reveal to students the logic and power of statistical inference and quickly focus on the holistic process of a statistical investigation, directly combating the issues described in Section 2 by finding the balance between 'proof' and 'mistrust.' In Section 4 we will go on to argue that these approaches should not only be a goal of 'new' introductory, applied statistics courses for non-majors, but should also be a primary directive for all undergraduate statistics courses. Finally, in Section 5, we lay out some broad next steps we see as necessary to achieve the goal of better statistical thinking throughout the undergraduate statistics curriculum, as catalyzed by more fully leveraging simulation-based inference.

**2. Anti-statistical thinking in traditional statistics courses**

There is an increasing societal need for data to inform decision making: no longer is it sufficient to make decisions based merely on intuition. This trend is now pervasive across disciplines and market sectors (Manyika et al., 2011). With this increased societal emphasis, statistical thinking has now moved to the forefront of daily life. Statistical thinking has been described as the need to understand data, the importance of data production, the omnipresence of variability, and the quantification and explanation of variability (Cobb, 1992; Hoerl & Snee, 2012; Snee, 1993; Wild & Pfannkuch, 1999). However, most students in introductory statistics courses fail to develop the statistical thinking needed to utilize data effectively in decision making (del Mas, Garfield, Ooms, & Chance, 2007; Martin, 2003). In a macro-sense, students tend to enter and leave most introductory statistics courses thinking of statistics in one of at least two incorrect ways:

1. Students believe that statistics and mathematics are similar in that statistical problems have a single correct answer; an answer that tells us indisputable facts about the world we live in (Bog #1: *overconfidence*) (Nicholson & Darnton, 2003; Pfannkuch & Brown, 1996), or,

2. Students believe that statistics can be 'made to say anything,' like 'magic,' and so cannot be trusted. Thus, statistics is viewed as disconnected and useless for scientific research and society (Bog #2: *disbelief*) (Martin, 2003; Pfannkuch & Brown, 1996).

Figure 1 illustrates this dichotomy. The tendency is for students to get stuck in one of the two bogs of anti-statistical thinking instead of appropriately viewing statistical thinking as a primary tool to inform decision making. This black-and-white view of the world of statistics is common when first learning a new subject area, and reflects a tendency to focus on lower-order learning objectives in introductory courses (e.g., knowledge; comprehension) (Bush, Daddysman, & Charnigo, 2014).

These broad, wrong-minded, 'take home messages' have been documented in different settings. For example, students who incorrectly conclude that the accuracy of the data depends solely on the size of the sample have failed to account for the impact of sample acquisition on potential bias in estimates (Bezzina & Saunders, 2014). Students who have this misconception are tending towards *overconfidence*, thinking that statistics is trying to provide a single correct answer (e.g., the underlying parameter value) and bigger samples always get closer to the true underlying parameter value. However, when trying to address this misconception, we have observed that statistics educators may have a tendency to show many examples of how biased sampling, question wording, question order, and a variety of other possible sampling and measurement issues can impact results in a dramatic way which can potentially lead students to believe that statistics are so sensitive to these issues that it is rare that results can be trusted (*disbelief*). For a recent example, see Watkins, Bargagliotti, & Franklin (2014) who suggest that over concern about small sample conditions can contribute to student distrust of statistical inference.

Misconceptions also exist in significance testing. Students often have a misconception that a p-value less than 0.05 means that the null hypothesis is wrong, failing to account for the possibility of a type I error (*overconfidence* (del Mas et al., 2007)). However, when teaching about type I errors, we've observed that students may be quick to latch onto the idea that 'we never know for sure' and wonder about the value of statistics in informing our understanding of populations, processes, and experimental interventions (*disbelief*). See Pfannkuch & Brown (1996) for further discussion about the tension between deterministic thinking (overconfidence) and statistical thinking.

Though less well studied, these misconceptions may not be limited to the introductory course, but could be pervasive throughout the undergraduate statistics curriculum. As little systematic research on the issue exists, we wonder how often statistics majors find themselves in situations without good statistical intuition. For example, the traditional probability and mathematical statistics sequence can emphasize applied calculus by focusing on how well a student can take integrals, with little to no opportunity for data analysis. Similarly, a course in regression may highlight having students perform matrix multiplication and partial derivatives, instead of spending time working with complex, real data sets. In both these examples, the focus is on the mathematical and deterministic aspects of statistics, which comes at the expense of the applied aspects, like the importance of sample acquisition. Evidence from the Stat 101 course suggests this approach can hinder students' ability to think statistically (del Mas et al., 2007; Pfannkuch & Brown, 1996). Should we expect less from more mathematically mature students who have little experience with data? If statistical thinking requires experience with data (DeVeaux & Velleman, 2008), then we argue that the advanced statistics curriculum should focus on data and reasoning.

We argue that addressing student misconceptions about statistical reasoning and practice and avoiding the two common ways of thinking 'anti-statistically' requires at least two broad curricular themes: (1) Students need to realize that statistical thinking is radically different than mathematical thinking (moving out of the bog of overconfidence), and (2) Students need to see statistics as quantitative support for the entire research process (moving out of the bog of disbelief). Within the last few years, innovations in the Stat 101 course, facilitated in large part through the use of simulation-based inference methods, directly address these two curricular themes. In the following section we will discuss the potential impact of simulation-based inference methods on the entirety of the undergraduate statistics curriculum in light of recent success in using these methods in Stat 101.

**Figure 1. Student and societal tendencies with regards to statistical thinking**

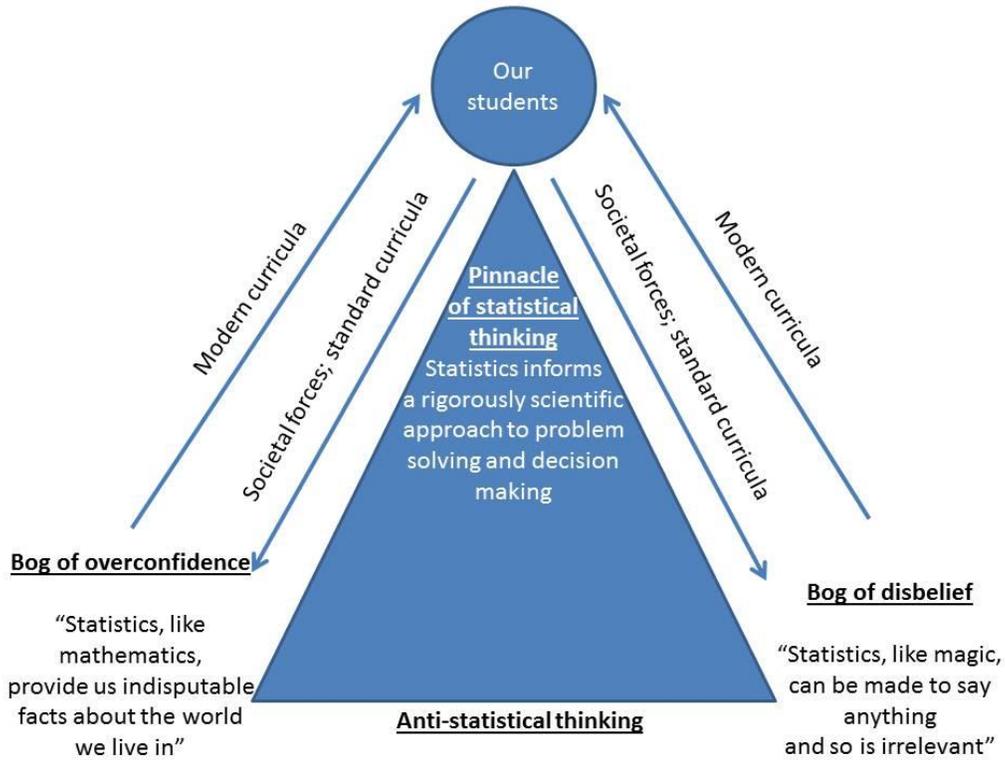

*Caption:* Like a ball at the top of a steep incline, students and society have a tendency to quickly fall into one of two bogs of 'anti-statistical thinking' leading to a view that statistics is irrelevant to science and society. A curriculum that focuses too much on lower-order learning objectives, like formulas and algebraic manipulation, may lead to overconfidence and a curriculum isolated from the entire research process may lead to disbelief. Curricula that emphasize higher order learning objectives--the scientific method and the entire statistical process from hypothesis formulation through communication of results-- may help students attain better statistical thinking.

**3. Impact of simulation-based approaches**

A major reform effort in Stat 101 courses involves the emphasis of simulation-based inference methods resulting in some recent curriculum which embrace this approach (Diez, Barr, & Cetinkaya-Rundel, 2014; Lock, Lock, Lock, Lock, & Lock, 2012; Tabor & Franklin, 2012; Tintle et al., 2015). A few examples of the use of simulation-based methods in the introductory course include:

(1) Simulating null distributions for a test of a single proportion using coins and spinners (e.g., Is 14 out of 16 correct choices out of two options unlikely to occur by chance alone?),

(2) Generating confidence intervals using (a) the bootstrap, (b) inversions of the test of significance and/or (c) estimated standard errors from simulated null distributions, and

(3) Simulating null distributions for two variable inference using permutation of the response variable.

Now that textbooks exist with this focus, more high school and university students experiencing these methods in their first algebra-based statistics course.  The authors of these textbooks claim that these methods can give students a deeper understanding of the reasoning of statistical inference and of the statistical investigation process as a whole.  Preliminary evidence supports comparable, if not improved, performance on validated assessment items (Tintle et al., 2014; Tintle, Topliff, VanderStoep, Holmes, & Swanson, 2012; Tintle, VanderStoep, Holmes, Quisenberry, & Swanson, 2011) and comparable student attitudes (Swanson, VanderStoep, & Tintle, 2014).We organize our summary of the potential benefits of this approach into three sections: (3.1) statistical over mathematical thinking, (3.2) highlight the entire research process and (3.3) good pedagogy.

**3.1. Simulation-based inference emphasizes statistical thinking over mathematical thinking**

Simulation-based inference offers at least three main innovations that emphasize statistical thinking over mathematical thinking, arguably moving students out of the bog of overconfidence. First, simulation-based inference does not rely on a formal discussion of probability before getting to the concepts of statistical inference. So, we can talk meaningfully with students  about the logic and scope of inference earlier in the course, and have students spend more time thinking critically about the sources and meaning of variability ( Cobb, 2007; Tintle et al., 2011). For example, students can do a coin tossing simulation to estimate a *p*-value for a test of a single proportion in the first week, if not the first day, of the course (Roy et al., 2014) with students able to answer whether "chance" is a plausible explanation for an observed sample majority. The earlier and more persistent discussion of these ideas is critically important in improving students' ability to think statistically. This discussion is further facilitated by curricular efficiencies often realized in courses utilizing simulation-based inference (e.g., efficient coverage of new inference situations, less time on abstract probability and sampling theory, potentially more efficiency with descriptive statistics (Tintle et al., 2011)) If statistics, as argued by DeVeaux and Velleman (2008), is like literature, then practice is critical to improving statistical thinking. Giving students more time to build their expertise and hone their statistical intuition by moving inference earlier in their education, facilitated by a simulation-based approach, is a critical step forward in improving students' understanding of the logic of inference, its place in a scientific investigation, and appreciation for the power and applicability of statistics. Notably, the Common Core State Standards in Mathematics (CCSSM) argue for introduction to basic simulation ideas in high school (NGACBP, 2010).

A second, but related, point is that simulation-based inference offers simpler and more intuitive choices and connections for students. For example, when first comparing two groups, the difference in group means (or proportions) is used. This means it is not necessary to compute the more complicated *t*-statistic or to even consider the mysterious degrees of freedom. Instead, *p*-values are always computed simply by counting how many simulated statistics are equal to or more extreme than the observed difference. This approach works for all sample sizes and comparisons: there is no need for students to focus their time memorizing which different buttons to push on the calculator or tables to use in an Appendix, which will depend on some arbitrary sample size condition cutoffs and/or choice of textbook. Simulation-based inference makes the logic of inference (e.g., compute the statistic, simulate the null hypothesis, and evaluate the strength of evidence (Tintle et al., 2015)) more prominent and requires less mathematical computation by students, meaning that the course can focus more on inductive, rather than deductive, reasoning.

Finally, simulation-based inference acts as a sandbox for students to explore more advanced statistical topics. For example, simulation-based methods have less need for large samples and symmetric distributions, and so are more widely applicable to real data. Furthermore, simulation-based inference methods are flexible and allow for student experimentation about summary statistics (e.g., What is a reasonable sample size to use before the null distribution behaves like a normal distribution? What factors does that depend on?) keeping students engaged in real, applied data analysis, while foreshadowing and potentially exploring ideas typically reserved only for upper-level undergraduate statistics students (e.g., What do you do if the validity conditions aren't met?).

**3.2. Simulation-based inference makes it easier to highlight the entire research process**

Simulation-based inference also helps support the idea that statistics involves the entire research process (arguably moving students out of the bog of disbelief). The Guidelines for Assessment and Instruction in Statistics Education (GAISE) College Report (GAISE College Group, 2005) list five parts of the statistical process through which statistics works to answer research questions: (1) How to obtain or generate data, (2) How to graph the data as a first step in analyzing data, and how to know when that's enough to answer the question of interest, (3) How to interpret numerical summaries and graphical displays of data- both to answer questions and to check conditions, (4) How to make appropriate use of statistical inference, and (5) How to communicate the results of a statistical analysis, including stating limitations and future work.

Because simulation-based methods make it possible to discuss inferential methods (confidence intervals and tests of significance) earlier in the course, students more readily have all of the tools needed to holistically focus on the entire research process, rather than be presented with the more "traditional" compartmentalized sequence of topics: descriptive statistics, data production, sampling distributions and, lastly, inference. For example, the *Introduction to Statistical Investigations* curriculum (Tintle et al., 2015) makes it a point to start the entire course by framing statistics in light of the "six steps of a statistical investigation" and having students consider all six steps in virtually every study presented throughout the book. Importantly, it ensures that connections between data production and analysis (Cobb, 2007) can be explored and help reinforce student learning. Students are also regularly considering limitations of studies (e.g., Does this answer the original research questions? What generalizations are allowed?) and practicing statistical communication. Furthermore, the presentation and exploration of new statistical techniques are always motivated by a genuine research study (and how the different type of study impacts all six steps), so that students start and end by thinking about a research question, not a mathematical 'what if?' question or seemingly unrelated computation.

**3.3. Simulation-based inference reflects good pedagogy**

Two important aspects of the GAISE College guidelines, as well as good pedagogy in any course (Freeman et al., 2014), are (a) the use of active learning, and (b) the use of assessment to drive student learning. Simulation-based inference is naturally conducive to an active learning approach in the classroom. Students frequently use tactile simulations and pool class results together, as well as discussing why some students may get different results and conclusions than others. These tactile methods are then mirrored and extended using technology in a way that aids visualization and intuition.

Furthermore, the use of simulation-based methods does not in any way preclude the full integration of other best practices that help students experience the fullness of statistics and the entire research process. For example, statistics courses should include a student-directed applied statistics research project which provides students the opportunity to experience the scientific method first-hand (Halvorsen, 2010; Singer & Willett, 1990). It is possible, however, to fail to fully realize all of the potential benefits of a project in an introductory course. But, because students in a simulation-based course often have a deeper and more integrated understanding of data collection, data exploration, and inference techniques earlier in the course, student projects have the potential to be deeper and richer student learning experiences. Furthermore, many traditional introductory statistics activities (Gelman & Glickman, 2000; Gnanadesikan, Scheaffer, Watkins, & Witmer, 1997) can still be a natural fit in courses using simulation-based methods.

Growing evidence suggests that the use of simulation-based methods does improve statistical thinking of introductory students. For example, Tintle et al. (2011) showed better post-course performance on the CAOS test (del Mas et al., 2007) with particular gains in areas related to inference and study design compared to the traditional curriculum. Furthermore, Tintle et al. (2012) demonstrated better retention of these concepts post-course. More recent evidence suggests good conceptual understanding (Chance & Mcgaughey, 2014; Tintle et al., 2014) and attitudes (Swanson et al., 2014) among students using simulation-based curricula at a variety of institutions, though additional data is needed to demonstrate that this performance is better than standard curricula. Maurer and Lock (2015) found improvement on questions related to confidence intervals in an introductory course utilizing the bootstrap compared to a course using a traditional asymptotic approach. Finally, recent evidence among students in a first statistics course showed improved understanding of inference after exposure to randomization methods, after being first exposed to traditional approaches (Case, Battle, & Jacobbe, 2014).

**4. Maximizing the impact of simulation-based inference methods throughout the undergraduate statistics curriculum**

Despite promising experimentation in Stat 101, we contend that the potential impact of simulation-based methods as a key component of improved statistical thinking has not been realized in the broader scope of the undergraduate statistics curriculum. Students who are more quantitatively trained (e.g., prior training in statistics and/or calculus) tend to see more of the mathematical foundations of statistics in their first course and/or throughout the curriculum. This connection to mathematical foundations is often at the expense of growing students' ability to think statistically, potentially leading to even statistics majors and minors ending up in the bogs of anti-statistical thinking, or failing to recruit more students into the major and minor to begin with. The success of the use of simulation-based methods in the algebra-based introductory statistics course can and should act as a catalyst for the development of alternative, simulation-based, introductory statistics courses for students with stronger mathematical backgrounds, as well as advanced classes (for both majors and non-majors) that continue to connect to and promote the simulation-based methods students see in their first course. In fact, we argue that simulation-based methods and the thinking they promote (e.g., could this result have happened just by chance?) should be a pervasive theme throughout the undergraduate statistics curriculum. However, currently, mathematically inclined students often fail to see these benefits and the logic and reasoning developed in the introductory course is not built upon in subsequent courses. In the following two sections we discuss introductory statistics for quantitative majors and courses beyond the introductory course. We then highlight common principles about the impact of simulation-based inference on courses throughout the undergraduate statistics curriculum.

*Introductory Statistics for Quantitative Majors*

Mathematically inclined students (e.g., those who have taken calculus or AP statistics) typically will not take Stat 101, but rather begin with a calculus-based introductory statistics or a probability-mathematical statistics sequence. Thus, these more mathematically mature students tend to miss out on the benefits of learning simulation-based inference, including more holistic thinking about how to conduct inference in any study and improved ability to reason about variability, among others. Notable exceptions include *Introduction to Statistical Concepts, Applications, and Methods,* an introductory statistics course for more mathematically inclined students which centers around simulation and the investigative process as a whole (Chance & Rossman, 2015) and an accelerated Stat 101 course (Tintle et al., 2013).

*Courses Beyond the Introductory Course*

Recent curricular efforts have developed materials for post-introductory students who have not had calculus, so that students can take a second or third applied course immediately after their first applied course in statistics (Cannon et al., 2013; Kuiper & Sklar, 2013; Legler & Roback, 2015; Ramsey & Schafer, 2013; Tintle et al., 2013). These courses have demonstrated that calculus is not needed before more advanced statistics courses are possible. With one exception (Tintle et al., 2013) these courses do not necessarily build on the simulation-based introductory statistics course.

*Common Principles*

When considering the potential impact of simulation-based inference on courses throughout the undergraduate statistics curriculum, there are four main principles that should be kept in mind---three principles that mirror lessons learned for Stat 101 (see Section 3), and one new principle that is directly relevant for all undergraduate statistics students.

*Principle #1. Emphasize statistical thinking over mathematical thinking*

In section 3.1, we made three main arguments as to how simulation-based inference can enhance statistical thinking over mathematical thinking, and help students move out of the bog of overconfidence. These same arguments apply to other courses for majors and minors as well. As students move to consider more advanced methods, it is critical to maintain the focus on the overall logic, and to give them a sandbox through which to explore these topics.

For example, simulation-based inference allows instructors to spend more time on the connections between test statistic choice, resulting null distributional shape and, ultimately, statistical power: Simulations of null distributions for the same dataset can be readily generated to allow for the comparison of strength of evidence using different test statistics. Students can then be asked to explain why certain test statistics provide stronger evidence than others for a particular dataset (e.g., difference between largest mean and smallest mean vs. *F*-statistic when comparing multiple group means). As time permits, these discussions can translate to more generalized discussions of the relationship between test statistic choice and statistical power.

Another example is the discussion of more complex study designs. For example, blocking can be presented as a study design to control excess variability. Students can be asked to develop a null-hypothesis simulation model for a simple blocked design (e.g., two treatments, and a single dichotomous blocking variable). Anecdotally, we have found that students can easily translate their understanding of null hypothesis simulations to this more complex design (namely, re-randomize treatments within the blocks) by applying the principle of matching the study design to the null hypothesis simulation strategy. We have used this as a launching pad into discussion of the analytic control of confounding variables in observational studies, acting as a bridge to general linear models, multiple regression, ANCOVA, etc.

Another example is having students apply different simulation models to the analysis of the relationship between two quantitative variables (e.g., regression slope). Some simulation-based inference curricula simulate the null hypothesis of no relationship between variables *x* and *y* by shuffling the values of the response variable. With the tools and intuition students have developed, it is straightforward to discuss alternative strategies including (1) taking the observed vales of *x*, the sample standard deviation of *y* given *x*, (*s*), and simulate normally distributed *y* values at each *x* using the hypothesized slope, or (2) simulating values from a bivariate normal distribution centered at $\bar{x}$, $\bar{y}$ and using $s_x$, *s* and the hypothesized slope, among others. Students can then consider how the standard error of the null distribution varies across the different model assumptions and why.

Even greater opportunities for exploration exist with more mathematically mature students, providing an opportunity to deepen and enrich student's statistical thinking by connecting to more sophisticated probability theory. These connections can be made by motivating probability theory via statistical questions: observing distributional patterns via simulation and then moving to more formal articulation of probabilistic phenomenon. For example, a permutation test comparing two independent groups on a binary response generates a null distribution that can be modelled exactly using Fisher's Exact Test. Thus, there is a natural segue into a discussion of the hypergeometric distribution and summing the tails of the resulting density function to yield *p*-values. We feel that when presenting this topic, students can be kept focused on statistical concepts by discussing issues like computational challenges in getting exact *p*-values for Fisher's exact test vs. increasing the number of permutations, finite population vs. process sampling as motivation for the binomial approximation to the hypergeometric distribution, and how best to define a two-sided p-value for an asymmetric distribution, instead of on the derivation of the hypergeometric density. Similar examples hold for continuous distributions, and serve to strengthen and maintain the ties between calculus and statistics (e.g., integration to find *p*-values). Another important point is that simulation provides answer questions not easily answerable by probability theory (e.g., the null distribution of the difference in percentiles between two or more independent groups). Thus, students can end up more empowered to answer sophisticated mathematical and statistical questions by starting with a simulation approach, and then be motivated to explore theoretical properties. Pfannkuch & Brown 1996) note that 'a formal mathematical approach to teaching probability may serve as an obstacle to the development of statistical thinking' We believe that simulation-based inference techniques can help walk this fine line: enhancing the statistical thinking of students, while not sacrificing important connections between mathematics and statistics. This approach has the benefit of contextualizing probability theory and providing students enhanced statistical and probabilistic intuition before shifting to theoretical concepts. For discussion about the use of simulation-based methods in teaching mathematical statistics see (Cobb 2011), for an example see Chihara & Hesterberg (2011).

*Principle #2. Emphasize the entire research process*

The arguments presented in section 3.2 (e.g., immediate and continual focus on the entirety of the research process to help address the bog of disbelief) also hold for an introductory course for more mathematically mature students, as well as for subsequent courses. With efficiencies gained by a simulation-based approach, students can spend more time considering issues of design and data integrity, messy data and data ethics, as well as spending more time reading and discussing applied scientific literature and use of statistics in the popular press. These projects and assignments force students to directly confront the application of challenging statistical ideas in a real situation. Simulation-based inference acts a catalyst for such projects by (a) providing curricular efficiencies (see 3.1) which free up more time for such projects, (b) providing a more flexible approach for data analysis when data don't fit typical statistical data assumptions and models, (c) allowing students to think about deeper, overarching statistical themes, by building on the stronger conceptual foundation arguably created by simulation-based inference (see 3.3).

Additionally, such courses and chapters within those courses can be built around types of research questions instead of mathematical theorems (Malone, Gabrosek, Curtiss, & Race, 2010), and underscore the importance of sample acquisition and study design (Ramsey & Schafer, 2013). Statistics courses should not be divorced from real data analysis, failing to contextualize the quantitative techniques being explored in the larger scientific framework. Students can and should be making these connections throughout their curriculum.

Finally, we note that capstone experiences within the major will benefit from students with a strong foundation in simulation based inference by (a) allowing for more innovation in the development of novel test statistics (Roback, Chance, Legler, & Moore, 2006) to deal with novel data situations, (b) acting as culmination of many research oriented experiences for students, and (c) giving students experience with statistics for what it really is early and often in their academic career.

*Principle #3. Make use of good pedagogical practices*

Finally, best-practices pedagogy should be kept at the forefront of the curricular design in all statistics courses. As noted earlier (section 3.3), simulation-based inference is naturally connected to active-learning classroom techniques (Freeman et al. 2014), and may enhance project experiences as well. Students' ability to explain the reasoning behind (all) statistical methods should also be much improved.

Perhaps the audience most benefited by this shift in focus is future teachers. Since teachers "teach how they were taught," (Bishop, Clements, Keitel, Kilpatrick, & Koon-Shing Leung, 2003) these ideas and reasoning process are much more consistent with current NCTM Standards, GAISE K-12 and College Guidelines, and CCSSM. Future teachers may be among those who benefit most from this shift in focus.

An additional critically important step is to train existing K-12 and college level teachers to teach statistics as a way of thinking about data that is separate from mathematics. A substantial challenge for the statistics education community is to train individuals who are mathematically inclined to teach statistical thinking. Simulation-based methods and experimental new courses are driving us to embrace statistical thinking---we need instructors to understand these approaches, why they are important, and how to encourage statistical thinking in the classroom and throughout the curriculum.

*Principle #4. Increasing practical relevance of simulation-based inference*

It is also worth noting that as students advance in the undergraduate statistics curriculum, it is increasingly important to recognize that simulation-based methods (including the bootstrap and permutation tests) are now becoming mainstream methods in the applied statistician's toolbox. Undergraduate curricula that provide early exposure to these modern, computationally intensive statistical methods, allow for deeper and broader thinking about these modern techniques later in the curriculum.

As students' progress in subsequent courses, they can be held more responsible for the design and implementation of simulation methods, giving them more of the computational skills increasingly in demand in today's world (ASA Workgroup, 2014).

**5. Next steps**

The statistical profession is at a crossroads. Academia, society, and industry are demanding more and more statistical literacy, increasingly counting on data to make decisions. This means that citizens must have basic statistical literacy and that more and more data experts are needed to satisfy the insatiable demand for data-informed decision making (Manyika et al., 2011). For over a century, statisticians have been the primary individuals meeting this need. However, shortsightedly, statisticians have kept their curriculum and courses tied to mathematics, simultaneously turning off generations of future statisticians to the profession and reinforcing misconceptions about statistics. Despite this substantial hurdle, refocusing the entire undergraduate statistics curriculum towards an emphasis on statistical thinking, an effort greatly enhanced through the use of simulation-based inference, may counteract the misconceptions about, and shortage of, statistically literate and trained individuals. We argue that simulation-based inference can act as a catalyst towards improved statistical thinking throughout the undergraduate statistics curriculum by (a) making clear the primary logic of inference and understanding of variability, (b) strengthening ties between statistics and probability/mathematical concepts, (c) encouraging a focus on the entire research process, (d) facilitating student thinking about advanced statistical concepts, (e) allowing more time to explore, do and talk about real research and messy data, and (f) acting as a solid foundation on which to build statistical intuition.

How then should we proceed? To take advantage of this unprecedented opportunity, we propose four recommendations for the statistics profession.

*Recommendation #1. We must be willing to break out of a 'mathematical' mindset in our courses and programs.* Most professional statisticians were mathematicians first, many statistics courses are taught by mathematicians and many statisticians live within mathematics departments. Furthermore, historically, the entirety of a student's traditional K-12 quantitative training focused on deductive reasoning with an eye towards a pinnacle of Calculus (Benjamin, 2009). This trajectory is present today both in the broad structure of the undergraduate curriculum, as well as in the content and pedagogy of many statistics courses. As we consider radical overhauls to our courses and our pedagogy we must consistently ask

"How will this promote statistical thinking and avoid the bogs?" Just because some students can do interesting mathematics, doesn't mean that they should do interesting mathematics in statistics courses. We also should not presume that mathematically mature students, even those in our statistics program, have necessarily broken out of the bogs of statistical thinking. Where in our curricula are they given that opportunity? Holistic statistical thinking is an art-form. Practice is needed to become an expert. Simulation-based inference acts as a catalyst for improving student's ability to think holistically.

*Recommendation #2. We must be willing to radically experiment with new courses and sequences in the undergraduate statistics curriculum.* The impact of a simulation-based first course early in the undergraduate curriculum goes even further than content carry-over and a stronger conceptual foundation on which to build. Because a simulation-based first course better reflects statistical thinking, and not mathematical thinking, and reflects recommended teaching practices, it may be more appealing to a broader and less mathematically mature set of students. Impacting even a small percentage of the large number of students who take statistics courses each (nearly 500,000 at U.S. four year and two-year colleges and universities alone in 2010 (Blair, Kirkman, & Maxwell, 2010)) to take additional coursework in statistics would have a huge impact on our society's statistical literacy. We must recognize that society is changing-- embracing data-informed decision making--, and so is K-12 education --promoting more statistical exposure. We must also embrace new ideas, as a way to have a chance to ride the big data wave and promote statistical thinking at the undergraduate level. In this paper, we have only presented the tip of the iceberg with regards to ideas for promoting statistical thinking in the undergraduate statistics curriculum. Much more thinking and discussion is needed. To facilitate those discussions there is a need for conference sessions, journal articles, email listservs and online forums (one example is http://www.causeweb.org/sbi), free sharing of materials and ideas, and more.

*Recommendation #3. We must better understand what students do and don't learn in our statistics courses and use these assessments to drive curricular change.* Finally, as we experiment and try new approaches, we must use assessment to drive curricular change. Unfortunately, few standardized

assessments of student thinking in the undergraduate statistics curriculum exist beyond the algebra-based introductory statistics course. Furthermore, assessments that focus on measures of statistical thinking in the introductory course tend to show relatively poor student performance. The development and utilization of assessments of statistical thinking are needed to drive continuous improvement of courses and curricula and further expose the gaps in current courses and programs.

*Recommendation #4. Embrace computation*. Simulation-based methods are computationally intensive approaches which often provide greater flexibility in their implementation as compared to traditional, asymptotic approaches. These methods are also growing in popularity in practice. Thus, simulation-based methods are well in line with recent curricular guidelines suggesting that we must embrace computation in our courses and curricula (ASA Workgroup, 2014).

## 6. Conclusions

Despite the significant work that is needed, we remain optimistic. The momentum behind the use of simulation-based methods in introductory courses, the endorsement of modern pedagogical standards for statistics teaching (GAISE College Group, 2005) and the development of the recent undergraduate program guidelines (ASA Workgroup, 2014) are all extremely encouraging signs of progress. Whereas the pessimist might point out that some of these arguments and approaches have been around for decades (CATS, 1994; Cobb & Moore, 1997; Cobb, 1992, 1993; Higgins, 1999; Waldrop, 1994), the persistent nature of these arguments and approaches keeps us encouraged. The integration of simulation-based methods opens the door to a variety of innovative pedagogy and content, enhancing students' abilities to think statistically. Our positive outlook about the opportunities before the statistics community is perhaps best summarized by a quote from the new 'front door' of the statistical discipline (American Statistical Association, 2014). Here you can read that "statistics is a science. It involves asking questions about the world and finding answers to them in a scientific way." If we can shape courses, curricula and instructors that embrace this philosophy the future of statistics is bright indeed.